\documentclass[twocolumn]{aastex631}

\usepackage{amsmath}

\shorttitle{High-Precision Galaxy Clustering}
\shortauthors{Doytcheva, Gerou, and Lange}

\newcommand{\hmpc}{h^{-1} \, \mathrm{Mpc}}
\newcommand{\hmsun}{h^{-1} \, M_\odot}

\begin{document}

\title{High-Precision Galaxy Clustering Predictions from Small-Volume Hydrodynamical Simulations via Control Variates}

\correspondingauthor{Alexandra Doytcheva}
\email{adoytch@umich.edu}

\author{Alexandra Doytcheva}
\affil{Department of Physics, University of Michigan, Ann Arbor, MI 48109, USA}

\author{Filomela V. Gerou}
\affil{Department of Physics, University of Michigan, Ann Arbor, MI 48109, USA}

\author[0000-0002-2450-1366]{Johannes U. Lange}
\affil{Department of Physics, University of Michigan, Ann Arbor, MI 48109, USA}
\affil{Leinweber Center for Theoretical Physics, University of Michigan, Ann Arbor, MI 48109, USA}
\affil{Department of Physics, American University, Washington, DC 20016, USA}

\begin{abstract}
    Cosmological simulations of galaxy formation are an invaluable tool for understanding galaxy formation and its impact on cosmological parameter inference from large-scale structure. However, their high computational cost is a significant obstacle for running simulations that probe cosmological volumes comparable to those analyzed by contemporary large-scale structure experiments. In this work, we explore the possibility of obtaining high-precision galaxy clustering predictions from small-volume hydrodynamical simulations such as MilleniumTNG and FLAMINGO via control variates. In this approach, the hydrodynamical full-physics simulation is paired with a matched low-resolution gravity-only simulation. By learning the galaxy--halo connection from the hydrodynamical simulation and applying it to the gravity-only counterpart, one obtains a galaxy population that closely mimics the one in the more expensive simulation. One can then construct an estimator of galaxy clustering that combines the clustering amplitudes in the small-volume hydrodynamical and gravity-only simulations with clustering amplitudes in a large-volume gravity-only simulation. Depending on the galaxy sample, clustering statistic, and scale, this galaxy clustering estimator can have an effective volume of up to around $100$ times the volume of the original hydrodynamical simulation in the non-linear regime. With this approach, we can construct galaxy clustering predictions from existing simulations that are precise enough for mock analyses of next-generation large-scale structure surveys such as the Dark Energy Spectroscopic Instrument and the Legacy Survey of Space and Time.
\end{abstract}


\section{Introduction}

Cosmological computer simulations are a key tool for our understanding of the Universe and its evolution from the Big Bang to today. Since the first such simulation performed by \cite{Press1974_ApJ_187_425} half a century ago, the complexity of these simulations has grown exponentially by many orders of magnitude. For example, while \cite{Press1974_ApJ_187_425} used $10^3$ tracer particles for the matter content of the Universe, state-of-the-art simulations use up to $\sim 10^{12}$ particles to probe cosmic structure formation \citep[see, e.g.,][]{Maksimova2021_MNRAS_508_4017, Ishiyama2021_MNRAS_506_4210, HernandezAguayo2023_MNRAS_524_2556, Heitmann2024_arXiv_2406_7276}. Similarly, whereas earlier simulations only calculated the forces of gravity, contemporary ``full-physics'' hydrodynamical simulations \citep[see, e.g.,][]{Vogelsberger2014_MNRAS_444_1518, Schaye2015_MNRAS_446_521} include all physical processes deemed relevant for the formation of galaxies such as hydrodynamics, star formation, supernova explosions, and active galactic nuclei.

Modern cosmology heavily relies on comparing the predictions from these simulations against observations of the Universe. Observational capabilities, in turn, also grow exponentially in size, reaching, for example, $\sim 10^{10}$ galaxies for the Rubin Observatory \citep{Ivezic2019_ApJ_873_111}. Ultimately, the volume of widely-used high-resolution full-physics simulations such as IllustrisTNG \citep{Nelson2019_ComAC_6_2}, $V \sim 0.03 \, \mathrm{Gpc}^3$ is significantly smaller than volumes probed in observations. For example, the Dark Energy Spectroscopic Instrument \citep[DESI; ][]{DESICollaboration2024_AJ_167_62} Bright Galaxy Survey \citep[BGS; ][]{Hahn2023_AJ_165_253}, currently probes a volume of $V \sim 1.7 \, \mathrm{Gpc}^3$ with a high density of galaxies \citep{DESICollaboration2024_arXiv_2404_3000}. Even the new generation of very large full-physics simulations such as MilleniumTNG \citep{HernandezAguayo2023_MNRAS_524_2556} and FLAMINGO \citep{Schaye2023_MNRAS_526_4978} still probe volumes significantly smaller than the DESI BGS year-1 data or sacrifice resolution substantially, preventing them to simulate all but high-mass galaxies. This mis-match in volume and resolution limits the precision and accuracy of simulation predictions and thereby comparisons against observational data. This problem has led to the development of several methods that aim to increase the predictive power of simulations, including the ``paired-and-fixed'' method \citep{Angulo2016_MNRAS_462_1} and the use of control variates \citep[CV; ][]{Chartier2021_MNRAS_503_1897, Kokron2022_JCAP_09_059, DeRose2023_JCAP_02_008, Hadzhiyska2023_OJAp_6_38, Ding2024_arXiv_2404_3117}.

Naively, one can estimate galaxy properties such as the galaxy stellar mass function and galaxy clustering by directly ``measuring'' these quantities in the full-physics simulation. However, this estimator will have significant noise due to the finite size of the simulation. Given that predictions of simulations should scale with the inverse square root of the simulated volume, improving predictive power of a given simulation is equivalent to increasing its volume. To achieve this, we use the CV statistical technique. Conceptually, this technique uses a cheap proxy simulation to estimate how the volume simulated in the full-physics run differs from the cosmic average. One then uses a modified estimator for the galaxy property in question that mitigates this difference. The CV technique is widely used in the context of cosmology, particularly for matter and galaxy clustering. However, existing applications are primarily focused on increasing the effective volume of gravity-only simulations via even cheaper proxy methods. In contrast, this work explores the potential of this technique to improve clustering predictions from full-physics simulations using gravity-only simulations plus a machine-learning model for the galaxy--halo connection as a proxy.

This paper is organized as follows. In section \ref{sec:data}, we briefly describe the simulation data we use. We outline our method for applying the control variates technique in section \ref{sec:methods}. We present our main results, including the increase in effective volume for galaxy clustering predictions, in section \ref{sec:results}. Finally, we discuss and summarize our results in sections \ref{sec:discussion} and \ref{sec:conclusion}, respectively. Throughout this work, $h = H_0 / (100 \, \mathrm{km} \, \mathrm{s}^{-1} \, \mathrm{Mpc}^{-1}) = 0.677$ is the reduced Hubble constant.

\section{Data}
\label{sec:data}

The IllustrisTNG project is a set of hydrodynamical simulations run with the AREPO code \citep{Weinberger2020_ApJS_248_32} that includes three simulation volumes, TNG50, TNG100, and TNG300, each of which are available with different mass resolutions. Compared to its predecessor, Illustris \citep{Nelson2015_AC_13_12}, IllustrisTNG has improvements such as including magnetic fields and improved kinetic AGN feedback and galactic wind models \citep{Weinberger2017_MNRAS_465_3291, Pillepich2018_MNRAS_473_4077}.

We use the redshift $z = 0$ snapshot of the highest-resolution simulation of the largest volume, TNG300-1, to analyze galaxy clustering and the galaxy--halo connection. TNG300-1 traces structure and galaxy formation in a cubic volume of side length $205 \, \hmpc$ using a particle mass of $7.6 \times 10^6 \, \hmsun$ and $4.0 \times 10^7 \, \hmsun$ for baryons and dark matter, respectively. From TNG300-1, we use all galaxies with a stellar mass ({\sc SubhaloStellarPhotometricsMassInRad}) above $10^{10} \, M_\odot$. In addition, to compute the lensing signal, we make use of a downsampled catalog of $\sim 7.4 \times 10^7$ particles. In addition to full-physics hydrodynamical simulations, the TNG300 volume also has gravity-only counterpart simulations, allowing us to learn the relation between gravity-only dark matter halo properties and galaxies in the full-physics run. Specifically, we use the TNG300-3-Dark version, the lowest-resolution gravity-only simulation of TNG300, that has a mass resolution of $3 \times 10^9 \hmsun$ for matter particles. From TNG300-3-Dark, we select all field halos above a total mass of $10^{11} \, \hmsun$, i.e., virtually all halos that may host a galaxy above $10^{10} \, M_\odot$ in stellar mass. Finally, a random $20 \%$ sub-sample of all particles is used to determine the gravitational lensing potential.

In this work, our goal is to predict the positions and abundances of galaxies in the low-resolution, gravity-only simulation, TNG300-3-Dark, through comparisons with the full-physics, high-resolution simulation TNG300-1. One approach would be to learn the relationship between galaxies in TNG300-1 and subhalos in TNG300-3-Dark. This approach would be advantageous since, in principle, it would only require us to determine for every subhalo whether it hosts a galaxy or not. Unfortunately, this approach is limited by the fact that certain galaxies and associated subhalos in TNG300-1 do not exist in TNG300-3-Dark due to tidal disruption caused by numerical resolution effects \citep{vandenBosch2018_MNRAS_474_3043} or the absence of stellar component that can prevent disruption \citep[see, e.g.][]{Vogelsberger2014_Natur_509_177}. Instead, we opt to associate galaxies to field halos identified with the friends-of-friends algorithm and learn the relationship between halo properties and the number of galaxies hosted. Galaxies in TNG300-1 are associated to halos in TNG300-3-Dark by searching for the nearest particle in TNG300-3-Dark that is associated with a halo. This way, every galaxy can be uniquely associated with a halo and $98.4\%$ of galaxies have a matching particle within $0.1 \, \hmpc$, i.e., a very secure association. We label galaxies as central galaxies if they are the most massive galaxy within $0.05 \hmpc$ of the most massive subhalo of the field halo in TNG300-3-Dark. If no galaxy is within $0.05 \hmpc$, the closest galaxy within $0.2 \hmpc$ is designated as the central. All other galaxies are labeled as satellite galaxies.

\section{Methods}
\label{sec:methods}

\subsection{Galaxy Clustering}
\label{subsec:clustering}

In this work, we analyze galaxy clustering, typically described via a series of summary statistics. One of the most commonly used summary statistics of galaxy clustering is the so-called galaxy two-point correlation function (2PCF). The 2PCF $\xi$ can be defined as the excess of galaxy pairs (so-called ``data-data pairs'', DD) separated by a distance $\vec{r}$, compared to the number of pairs expected for a fully random distribution (``random-random pairs'', RR),
\begin{equation}
    \begin{split}
    1 + \xi(\vec{r}) = \frac{\mathrm{DD}(\vec{r})}{\mathrm{RR}(\vec{r})} = \frac{\mathrm{DD}(\vec{r})}{N_{\rm gal} (N_{\rm gal} - 1) V_{\rm sim}^{-1} V_{\rm search}} \, .
    \end{split}
\end{equation}
In the last expression, we have written down the expected number of random-random pairs for a sample of $N_{\rm gal}$ galaxies in a periodic simulation box of volume $V_{\rm sim}$ when pairs are searched for in a volume $V_{\rm search}$ around each galaxy.

Due to the Universe's isotropy, if one has access to the actual three-dimensional galaxy coordinates, measuring the 2PCF as a function of $r = |\vec{r}|$ is a natural choice, i.e., measuring pairs in shells around each galaxy. This so-called real-space correlation function $\xi(r)$ is one of the galaxy clustering statistics we analyze here and independent of galaxy velocities. However, in actual observations, we typically do not have access to the intrinsic three-dimensional coordinates. Galaxy positions are inferred from redshifts which are impacted by the peculiar velocities of galaxies, so-called redshift-space distortions. If we assume that the line of sight of the observer is aligned with the $z$-axis of the simulation, then the apparent $z$-coordinates of galaxies at redshift $0$ will be modified according to
\begin{equation}
    z \rightarrow z + \frac{v_z}{H_0} = z + \frac{v_z}{100 \, \mathrm{km/s}} \, h^{-1} \, \mathrm{Mpc} \, ,
\end{equation}
while the $x$ and $y$-coordinates are unaffected. These new coordinates are called redshift-space coordinates and break the isotropy of the real-space coordinates. As a result, we quantify the 2PCF with respect to the separation $r_p$ perpendicular to the line of sight (the $x$ and $y$-coordinates) and the separation $r_\pi$ along the line of sight (the $z$-coordinate).

The so-called projected correlation function is defined via
\begin{equation}
    w_p (r_p) = \int_{-\pi_{\rm max}}^{\pi_{\rm max}} \xi(r_p, r_\pi) \mathrm{d} r_\pi \, ,
\end{equation}
where we choose $\pi_{\rm max} = 40 \, \hmpc$. The projected correlation function is intentionally designed to be mostly insensitive to redshift-space distortions by integrating along the $z$-axis. Conversely, the redshift-space multipole moments are explicitly designed to capture redshift-space distortions. They are defined via
\begin{equation}
    \xi_\ell(s) = \frac{2 \ell + 1}{2} \int\limits_{-1}^{+1} L_\ell(\mu) \xi(s, \mu) \mathrm{d} \mu \, ,
\end{equation}
where $s^2 = r_p^2 + r_\pi^2$ is the three-dimensional separation in redshift space, $\mu = r_\pi / s$ is the cosine of the angle with respect to the line of sight, and $L_\ell$ is the Legendre polynomial of order $\ell$. In this work, we will study the redshift-space multipole moments of order $\ell = 0$, $2$, and $4$, the so-called monopole, quadrupole and hexadecapole moments.

Finally, another common observable is galaxy--galaxy lensing whereby we measure gravitational lensing distortions of the shape of background ``source'' galaxies around foreground ``lens'' galaxies. This gravitational lensing signal is a measure of the galaxy--matter cross-correlation through the so-called excess surface density (ESD),
\begin{equation}
    \Delta\Sigma(r_p) = \frac{2}{r_p^2} \int\limits_0^{r_p} \tilde{r}_p \Sigma(\tilde{r}_p) \mathrm{d\tilde{r}_p} - \Sigma(r_p) \, ,
\end{equation}
where $\Sigma$ is the mean projected matter density along the $z$-axis.

We calculate the galaxy 2PCFs using the {\sc Corrfunc} package \citep{Sinha2020_MNRAS_491_3022} whereas the ESD is computed via {\sc halotools} \citep{Hearin2017_AJ_154_190} using the methods outlined in \cite{Lange2019_MNRAS_488_5771}.

\subsection{Control Variates}

Our work utilizes the CV method to compare clustering predictions from the full-physics simulation to a cheaper gravity-only plus machine-learning simulation of the same volume, as well as this cheaper model run on a greater volume. For any galaxy clustering property $\xi$, the CV estimate is
\begin{equation}
    \xi_{\rm CV} = \xi_{\rm FP, s} + c (\xi_{\rm GML, s} - \xi_{\rm GML, l}) \, ,
\end{equation}
where $\xi_{\rm FP, s}$ is the galaxy clustering in the (small) full-physics simulation and $\xi_{\rm GML,s}$ and $\xi_{\rm GML,l}$ are clustering properties calculated from the gravity-only simulations for the small and large volume, respectively. Finally, $c$ is a constant whose optimal value depends on the correlation between $\xi_{\rm FP, s}$ and $\xi_{\rm GML, s}$. Regardless of the choice of $c$, the CV estimator has the same expectation value as $\xi_{\rm FP, s}$, i.e., is an unbiased estimate of galaxy clustering in the full-physics simulation. This follows from $\langle c (\xi_{\rm GML, s} - \xi_{\rm GML, l}) \rangle = c (\langle \xi_{\rm GML, s} \rangle - \langle \xi_{\rm GML, l} \rangle) = 0$. However, for a suitable chosen value for $c$, i.e.,
\begin{equation}
    c = - \frac{\mathrm{Cov}(\xi_{\rm FP, s}, \xi_{\rm GML, s})}{\mathrm{Var}(\xi_{\rm GML, s}) + \mathrm{Var}(\xi_{\rm GML, l})}
\end{equation}
the variance is reduced by
\begin{equation}
    \frac{\mathrm{Var}(\xi_{\rm CV})}{\mathrm{Var}(\xi_{\rm FP, s})} = 1 - \rho^2 \frac{\mathrm{Var}(\xi_{\rm GML, s})}{\mathrm{Var}(\xi_{\rm GML, s}) + \mathrm{Var}(\xi_{\rm GML, l})} \, .
    \label{eq:reduction}
\end{equation}
where $\rho$ the Pearson correlation coefficient of the two estimates in the small-volume simulation. In the limit of the large-volume simulation being infinitely large or having infinitely many realizations of it, i.e., $\mathrm{Var}(\xi_{\rm GML, l}) = 0$, the variance reduction simplifies to $1 - \rho^2$. Ultimately, only the correlation coefficient $\rho$ matters for the variance reduction. As a result, in order to test the effectiveness of the CV method, we only need to study the small-volume simulations to measure $\rho$. In this work, we do not study the large-volume gravity-only simulation that would be needed in an actual application of the CV method.

It is worth re-emphasizing that even if $\xi_{\rm GML}$ is a biased estimate of $\xi_{\rm FP}$, the CV method will give an unbiased estimate of the clustering in the full-physics simulation. Additionally, no assumptions about the distribution of $\xi$ such as a Gaussian distribution was made.

\subsection{Galaxy Samples}
\label{subsec:samples}

\begin{figure}
    \centering
    \includegraphics{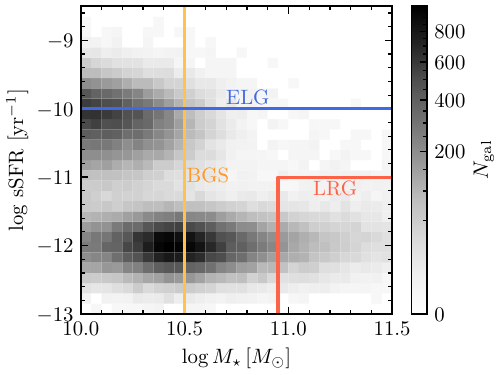}
    \caption{Definition of different galaxy samples in the plane of sSFR and stellar mass. The BGS sample is defined by a variable cut in stellar mass, as indicated by the orange line. LRG are indicated by the two red lines, where we isolated galaxies below a certain sSFR and above a stellar mass. The cut for ELGs is represented by the galaxies above the horizontal blue line. To make low-sSFR visible in this plot, the sSFR of every galaxy was increased by $10^{\mathcal{N}(-12, 0.25)} \, \mathrm{yr}^{-1}$, where $\mathcal{N}(-12, 0.25)$ denotes a random variable drawn from a Gaussian with mean $-12.0$ and scatter $0.25$.}
    \label{fig:samples}
\end{figure}

We define galaxy samples broadly designed to emulate the Luminous Red Galaxies \citep[LRGs; ][]{Zhou2023_AJ_165_58}, Emission Line Galaxies \citep[ELGs; ][]{Raichoor2023_AJ_165_126}, and Bright Galaxy Survey \citep[BGS; ][]{Hahn2023_AJ_165_253} samples in DESI. LRGs are old and massive galaxies that have ceased star formation and exhibit red spectral energy distributions as a result. To simulate the LRG sample, we select galaxies above a mass of $10^{10.95} \, M_\odot$ and below a specific star-formation rate (sSFR) of $10^{-11} \, \mathrm{yr}^{-1}$. This cut results in a number density $5 \times 10^{-4} \, h^3 \, \mathrm{Mpc}^{-3}$, similar to the LRG density in DESI. ELGs are typically irregular and late-type spiral galaxies. We define them via $\mathrm{sSFR} > 10^{-10} \, \mathrm{yr}^{-1}$, resulting in a number density of $1 \times 10^{-3} \, h^3 \, \mathrm{Mpc}^{-3}$, roughly mirroring the ELG density in DESI. The selection of galaxies from the BGS relies on an apparent $r$-band magnitude limit. As a result, the cut on the absolute $r$-band magnitude, which in turn strongly correlates with stellar mass, is highly redshift-dependent. To simulate this, we define three stellar-mass limited samples, BGS-1, BGS-2, and BGS-3, with number densities of $1$, $3$, and $9 \times 10^{-3} \, h^3 \, \mathrm{Mpc}^{-3}$, respectively. For reference, the corresponding stellar mass cuts are $\sim 10^{10.8}$, $\sim 10^{10.5}$, and $\sim 10^{10.0} \, M_\odot$. The different sample selections are illustrated in Fig.~\ref{fig:samples}. The different galaxy samples have vastly different clustering properties, number densities and galaxy--halo connections, allowing us to test our method across a representative range of galaxy samples.

\subsection{Galaxy--Halo Connection}

As described in section \ref{sec:data}, we can uniquely associate every galaxy in the full-physics simulation to a dark matter halo in the gravity-only counterpart. In particular, given a galaxy sample, we can determine the number of centrals and satellites for each halo in the gravity-only simulation. This data can be used as input for a supervised machine learning algorithm to learn the relation between halo properties and the number of centrals and satellites. Afterwards, the trained algorithm can be used to predict the number of galaxies for halos not in the training set.

We study the relation between galaxies and a variety of halo properties. These halo properties include the total halo mass, $M_{\rm tot}$, and the radius, $r_{\rm half}$, containing half the mass of the most-massive subhalo which acts as a proxy for halo concentration. Furthermore, we use the maximum circular velocity, $V_\mathrm{max}$, of the main subhalo, its spin, $\lambda$, and its particle velocity dispersion, $\sigma$. While $M_{\rm tot}$ is the mass of the entire halo, we also use $M_{\rm main}$, the mass of the most massive, i.e., the main subhalo, a proxy of the amount of substructure. Additionally, we use the redshifts $z_{10}$ and $z_{50}$ at which the main subhalo first attained $10\%$ and $50\%$ of its final mass as proxies for the halo formation history. Finally, the halo environment $\delta$ is measured by determining the mass density within a $5 \, \hmpc$ tophat filter.

In this work, our machine learning approach works as follows. First, input halo properties are normalized such that they follow a normal distribution with mean $0$ and unit variance. We then train a neural network to learn the relation between the normalized halo properties and the number of central and satellite galaxies. We use a fully connected network as implemented by the {\sc MLPRegressor} class in {\sc scikit-learn}. We optimized the hyper-parameters of the network using K-fold cross-validation. Ultimately, the network architecture we used has $3$ hidden layers with $128$, $64$, and $32$ neurons, respectively, and employs a $\tanh$ activation function. To avoid spurious over-fitting, we use a K-fold approach, training the network on a random $80 \%$ of halos and predicting number of galaxies for the remaining $20 \%$ of halos. This step is repeated $50$ times such that the number of centrals and satellites is predicted $10$ times for every halo. We then take the average of the $10$ neural network predictions for each halo.

After applying the machine learning algorithm, we have the expected number of centrals, $\hat{N}_{\rm cen}$, and satellites, $\hat{N}_{\rm sat}$, for all halos but not their positions or velocities. For centrals, we assign positions and velocities for galaxies based on the position and velocity of the main subhalo of each halo. For satellites, we instead pick the positions and velocities of $N_{\rm part}$ random particles associated with the halo. Thus, for each halo we have $N_{\rm part} + 1$ tracers for the positions and velocities of potential galaxies. The weight assigned to each tracer is $w = \hat{N}_{\rm cen}$ for centrals and $w = \hat{N}_{\rm sat} / N_{\rm part}$ for satellites. When computing galaxy clustering and lensing, data-data pairs are weighted by the product of the weights of the tracers. Similarly, for lensing, the lensing signal is calculated based on a weighted average of all tracers. This approach, using tracers with non-unity weights, allows us to significantly reduce shot noise in the clustering prediction of the machine learning algorithm.

\begin{figure}
    \centering
    \includegraphics{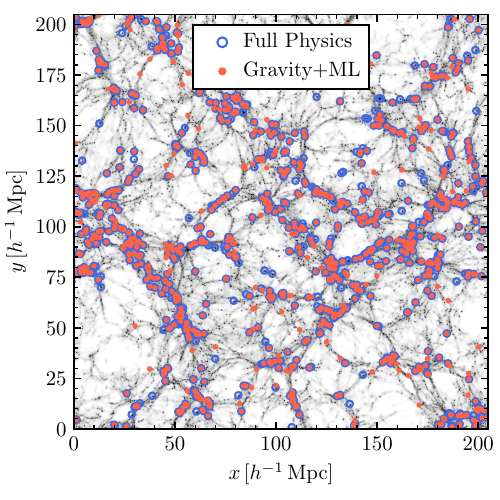}
    \caption{Large-scale structure in TNG300 at coordinates $z < 10 \, \hmpc$. The background color denotes the matter density in the TNG300-3 Dark simulation. Blue points denote the BGS-2 sample in TNG300-1 whereas red points are corresponding galaxies assigned to dark matter halos in TNG300-3 Dark by learning the galaxy--halo connection from TNG300-1.}
    \label{fig:visualization}
\end{figure}

Finally, to reduce computational cost, we downsample the original tracer sample by defining a minimum weight $w_{\rm min}$. Tracers with $w < w_{\rm min}$ are downsampled with a probability $w / w_{\rm min}$ and assigned a new weight $w / (w / w_{\rm min}) = w_{\rm min}$ whereas tracers with $w \geq w_{\rm min}$ are not downsampled. We choose $w_{\rm min} = 0.1$ for all calculations in this work, finding that further reductions in this value do not significantly alter the results in this work. Finally, we can use $w_{\rm min} = 1$, which results in all tracers representing one predicted galaxy, to visualize the predicted galaxy sample, as depicted in Fig.~\ref{fig:visualization}. For this example, we chose the BGS-2 sample using all available halo properties to train the neural networks. We see that our machine learning algorithm predicts a galaxy population that closely mimics the large-scale distribution of galaxies in the full-physics simulation.

\section{Results}
\label{sec:results}

In this section, we describe the main results of this work. We first describe how the prediction of the halo occupation depends on secondary halo properties besides halo mass. Next, we analyze how well our machine learning method reproduces galaxy clustering. Finally, we present the variance reduction our CV method can achieve.

\subsection{Halo Occupation}

\begin{figure*}
    \centering
    \includegraphics{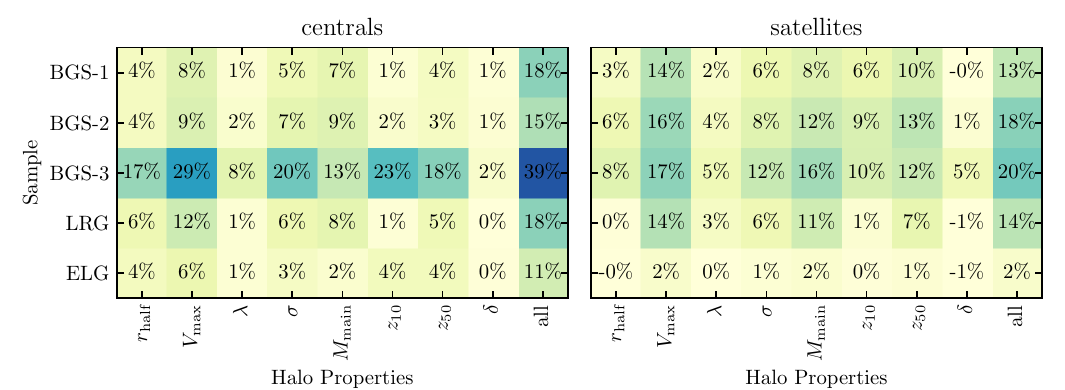}
    \caption{Visualization of how different secondary halo properties improve machine learning models of galaxy occupation over a mass-only approach. For each combination of galaxy sample and secondary halo property, we show $\Delta R^2$, the fractional reduction in the mean-squared error for the predicted galaxy occupation over a mass-only prediction. We separate by central galaxies (left panel) and satellite galaxies (right panel). The rightmost column in each panel denotes the inclusion of all available secondary halo properties.}
    \label{fig:r2}
\end{figure*}

To measure the performance of the galaxy--halo connection model, we use the coefficient of determination, $R^2$, between the actual, $n$, and predicted, $\hat{n}$, galaxy occupation,
\begin{equation}
    R^2 = 1 - \frac{\langle (n - \hat{n})^2 \rangle}{\langle (n - \langle n \rangle)^2 \rangle} \, .
\end{equation}
In the above equation, $\langle \rangle$ denotes the average over all halos. To look at the improvements in predictive power when incorporating halo properties beyond mass, we define
\begin{equation}
    \Delta R^2 = \frac{R_\mathrm{s}^2 - R_\mathrm{m}^2}{1 - R_\mathrm{m}^2} \, ,
\end{equation}
where $R_\mathrm{s}^2$ and $R_\mathrm{m}^2$ are the coefficients of determination for the model including secondary halo properties and the model with only halo mass, respectively. $\Delta R^2$ thus measures the fraction of the variance in the mass-only model that can be explained by secondary halo properties. In Fig.~\ref{fig:r2}, we show $\Delta R^2$ for different galaxy populations and secondary halo properties. We see that $V_\mathrm{max}$ and $M_{\rm main}$ are among the most important secondary halo properties to increase the predictive power. Contrary, the environment parameter $\delta$ is one of the least important parameters to predict occupation.\footnote{We note that some values for $\Delta R^2$ for environment are actually negative. This is likely due to overfitting when the underlying model depends on two properties, mass and environment, instead of one, halo mass.} Finally, in almost all cases, including all secondary halo parameters leads to a significantly stronger increase in predictive power than any individual halo property by itself. This demonstrates that our neural networks are able to model the complex, multi-dimensional interplay between halo properties and galaxy occupation. The improvement over the mass-only model is strongest for the high-density BGS sample, BGS-3, which has a high fraction star-forming galaxies, unlike BGS-1 and BGS-2.

\subsection{Galaxy Clustering}

\begin{figure}
    \centering
    \includegraphics{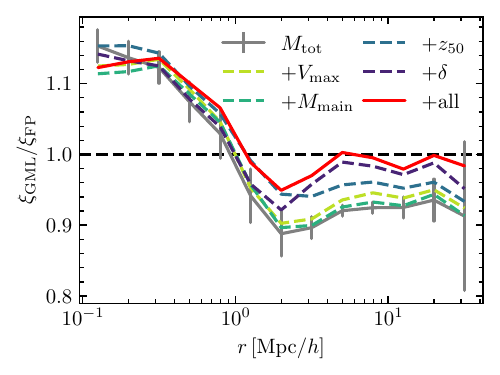}
    \caption{The ratio of the real-space correlation function $\xi$ for BGS-3 in the gravity-only plus machine learning simulation compared to the original full-physics simulation. The gray line indicates if halo mass alone, $M_{\rm tot}$, is used to learn the galaxy--halo connection. Furthermore, we show results if additionally the maximum circular velocity $V_\mathrm{max}$ (light green), main subhalo mass $M_{\rm main}$ (dark green), environment $\delta$ (turquoise), halo formation redshift $z_{50}$ (blue), or all available halo properties (red) are used. Error bars denote $68\%$ credible intervals and are derived from jackknife-resampling.}
    \label{fig:clustering}
\end{figure}

We now study how the gravity-only plus machine learning model predicts galaxy clustering. Fig.~\ref{fig:clustering} represents how different secondary halo properties affect real-space clustering predictions from the machine learning model for the high-density BGS sample, BGS-3. In all cases, on smaller scales, $r < 1 \, \hmpc$, the correlation function is overpredicted, likely indicating a mismatch in the halo profile of satellite galaxies. This could be because we assign satellite galaxies to random dark matter particles whereas \cite{McDonough2022_ApJ_933_161} found that satellites in the TNG model are actually slightly less concentrated than the dark matter. On the other hand, on larger scales for the mass-only model, the clustering amplitude is underpredicted. This mismatch in the large-scale galaxy bias for the mass-only model indicates the presence of galaxy assembly bias in IllustrisTNG \citep{Hadzhiyska2020_MNRAS_493_5506}, i.e., that galaxy occupation depends on secondary halo properties correlated with halo clustering. The ratio of large-scale clustering amplitudes does not change substantially when the main subhalo mass $M_{\rm main}$, maximum circular velocity $V_\mathrm{max}$, or halo formation redshift $z_{50}$ are used in addition to halo mass for the machine learning model. Similar conclusions hold for all secondary properties but the halo environment: including $\delta$ in the machine learning predictions, either in combination with only halo mass or all halo properties, brings the large-scale clustering predictions into good agreement with the full-physics simulation. We show here the results for BGS-3 due to the high number density and small uncertainties but results are similar for the other galaxy samples.

Interestingly, the secondary halo property most relevant for predicting galaxy clustering, environment $\delta$, is one of the least important properties for predicting galaxy occupation, as shown in Fig.~\ref{fig:r2}. As shown in \cite{Hadzhiyska2020_MNRAS_493_5506}, environment is also the most strongly correlated with large-scale bias. As a result, even though $\delta$ only reduces the mean-squared error for predicted galaxy occupation by $\sim 2\%$ for the BGS-3 sample, it has a significant impact on galaxy clustering. Reassuringly, Fig.~\ref{fig:r2} and~\ref{fig:clustering}, show the highest increase in accuracy and predictive power when all secondary properties are included. Finally, it is worth emphasizing that the CV method gives unbiased clustering estimates of the full-physics simulation regardless of the accuracy of the machine learning simulation. In other words, even using the mass-only model would lead to unbiased predictions on both small and large scales.

\subsection{Variance Reduction}

\begin{figure*}
    \centering
    \includegraphics{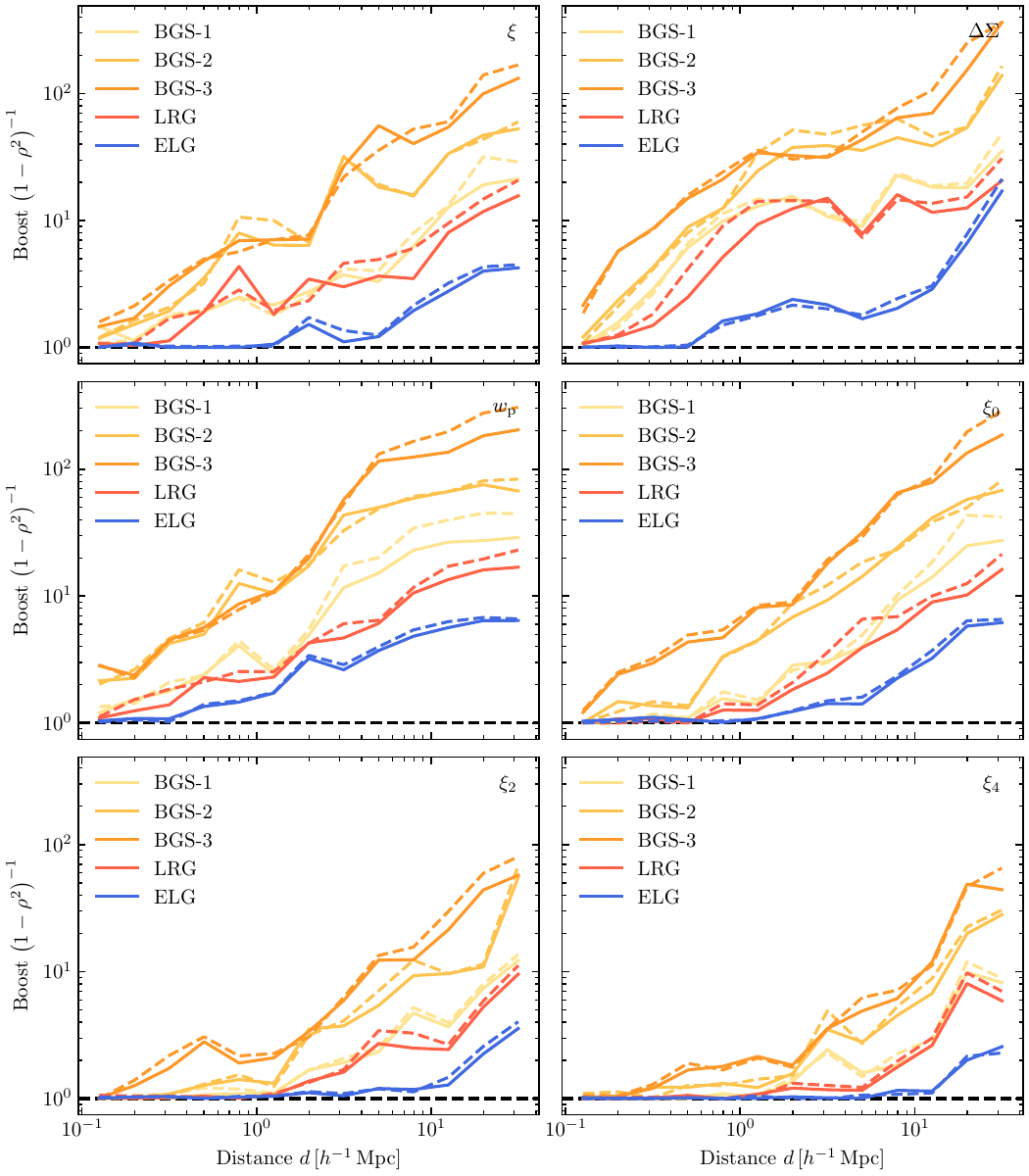}
    \caption{Effective volume increase for 3 different densities of our BGS sample, (BGS-1, -2, and -3) in ascending order of density, as well as ELG and LRG are shown as a function of distance for six clustering measures: real-space correlation function ($\xi$, top left), galaxy-galaxy lensing ($\Delta\Sigma$, top right), projected correlation function ($w_{\rm p}$, center left), redshift-space monopole ($\xi_0$, center right), quadrupole ($\xi_2$, bottom left), and hexadecapole ($\xi_4$, bottom right). The dashed lines indicate our ML method when only using halo mass to learn galaxy--halo connection, while the solid lines indicate that all halo properties were used.}
    \label{fig:boost}
\end{figure*}

The goal of this work is to develop a method to make highly precise clustering predictions from full-physics simulations. Through the use of CV, the variance of these predictions is reduced by up to a factor of $1 - \rho^2$, as described in eq.~\eqref{eq:reduction}, where $\rho$ is the Pearson cross-correlation coefficient between the machine-learning and the full-physics simulation. In the absence of CV, the variance should scale as inversely proportional to volume, such that the CV technique leads to an effective volume increase of up to $(1 - \rho^2)^{-1}$.

We measure $\rho$ through jackknife resampling of $4^3 = 64$ equal-volume parts of the simulation. As discussed in \citep{Mohammad2022_MNRAS_514_1289}, for auto-correlation functions, care has to be taken regarding the weight assigned to cross-pairs, i.e., pairs where one galaxy falls inside and one outside the jackknife region. Dy default, we use the v$_\mathrm{match}$ weighting method that was found to be highly accurate \citep{Mohammad2022_MNRAS_514_1289}. However, our results do not change significantly if we instead employ the widely-used v$_\mathrm{mult}$ weighting despite biases in the absolute variance \citep{Norberg2009_MNRAS_396_19}. Similarly, we did not find our results to be sensitive to choice of the number of jackknife regions, e.g., $27$, $64$, or $125$.

Fig.~\ref{fig:boost} represents the effective volume increase for six different clustering statistics described in section~\ref{subsec:clustering}: real-space correlation function $\xi$, galaxy-galaxy lensing $\Delta\Sigma$, projected correlation function $w_{\rm p}$, redshift-space monopole $\xi_0$, quadrupole $\xi_2$, and hexadecapole $\xi_4$. We see that the BGS samples typically have the largest effective volume increases for all clustering measures. As the density of the BGS samples increases so does their effective volume. The BGS-3 sample specifically, has a promising volume increase by a factor of up to $\sim 100$ for all clustering measures except for the hexadecapole, especially on large scales. LRGs have a volume increase by a factor of up to $\sim 10$, and ELGs are the least increased across all measures and scales, almost always staying well under a factor of $10$. It is important to note that the ELG and BGS-1 samples have similar number densities, $n \approx 10^{-3} \, h^3 \, \mathrm{Mpc}^{-3}$, indicating the galaxy density alone does not determine the volume increase.

We also find that across all clustering functions and galaxy samples, the variance reduction is greatest on large scales. Interestingly, when comparing the effective volume increase using only halo mass versus all halo properties, the difference is only of order $6\%$ when averaged across all statistics, scales, and samples. This difference does not demonstrate a significant improvement between the use of only halo mass and the use of all secondary properties for the purpose of variance reduction.

\section{Discussion}
\label{sec:discussion}

\subsection{Galaxy--Halo Connection}

While our work is primarily focused on variance reduction of galaxy clustering predictions, our analysis also focuses on modeling the galaxy--halo connection with machine learning \citep[see, e.g., ][and references therein]{Hausen2023_ApJ_945_122, Delgado2022_MNRAS_515_2733, Das2024_arXiv_2406_6103}. In this work, we focus on predicting the number of galaxies per halo as opposed to galaxy properties such as stellar mass and star formation rate, making our results closely comparable to the results of \cite{Delgado2022_MNRAS_515_2733}. The authors study a stellar mass-limited sample of galaxies that is in between our BGS-1 and BGS-2 samples in IllustrisTNG. Using a combination of random forest and symbolic regression, they find that the performance of the random forest in predicting the number of galaxies in each halo in their test set was very similar to the mass-only model, similar to our results as shown in Fig.~\ref{fig:r2}. In agreement with our findings, they also find that a mass-only model under-predicts large-scale clustering, an indication of galaxy assembly bias \citep{Hadzhiyska2020_MNRAS_493_5506}, and that environment is the most important secondary halo property to bring the clustering predictions of the machine learning model into agreement with the full-physics simulation.

Our analysis shows that the inclusion of all halo properties into the galaxy--halo connection does not substantially improve the galaxy occupation prediction and, as a result, the effectiveness of the CV method over a simply mass-only approach. One possibility is that the machine learning algorithm could be improved through a better architecture, more halo features, or simply more data coming from larger simulations. At the same time, galaxy formation is inherently chaotic. As demonstrated by \cite{Genel2019_ApJ_871_21}, minute differences in the initial conditions of full-physics simulation can substantially change the properties of $z=0$ galaxies whereas this so-called butterfly effect is much smaller in simulations without feedback, i.e., gravity-only simulations. Ultimately, this suggests that the properties of galaxies in the full-physics simulation cannot be perfectly predicted from the gravity-only counterpart, regardless of how many halo properties are considered. We leave investigations into more precise machine learning models of the galaxy--halo connection to future work, noting that such models may further improve the CV method presented here.

\subsection{Variance Reduction}

Given the computational costs, energy usage, and carbon footprint of large-scale cosmological simulations, variance reduction techniques are an active area of research. The use of CV, in particular, has been explored in a variety of previous works \citep[see, e.g.,][]{Chartier2021_MNRAS_503_1897, Kokron2022_JCAP_09_059, DeRose2023_JCAP_02_008, Hadzhiyska2023_OJAp_6_38, Ding2024_arXiv_2404_3117}. In these studies, the target ``expensive'' estimator is a gravity-only simulation and, possibly, an empirical galaxy model. The proxy estimator are analytic prescriptions, i.e., the Zel'dovich approximation, or approximate gravity solvers such as {\sc FastPM} \citep{Feng2016_MNRAS_463_2273} coupled with galaxy bias or halo occupation distribution (HOD) models. In contrast, in this work, the target estimator is a full-physics simulation and the proxy estimator a gravity-only simulation partnered with a machine learning model of the galaxy--halo connection. In principle, previously used proxy estimators could also be used for hydro-dynamical simulations as targets. However, given the very high computational cost of full-physics simulations\footnote{TNG300-1 full-physics simulation took $35$ million CPU hours to compute \citep{Nelson2018_MNRAS_475_624}. Contrary, TNG300-3-Dark only took $6$ thousand CPU hours to finish (D. Nelson, private communication).} compared to all proxy estimators, it makes sense to use the proxy with the largest variance reduction.

Comparisons with previous works is hampered by the fact that most focus on larger spatial scales than considered here as well as clustering properties in Fourier space, i.e., the galaxy power spectrum. \cite{Ding2024_arXiv_2404_3117} study the variance reduction for an DESI LRG-like sample constructed from a gravity-only simulation coupled with an HOD model where {\sc FastPM} simulations were used as proxy estimators. For the redshift-space monopole at a distance of $30 \hmpc$, they find a variance reduction of a factor of $\sim 4$ whereas we find a factor of $\sim 20$. At the same time, this comparison should be taken with a grain of salt given differences in the galaxy samples. For example, the LRG sample in \cite{Ding2024_arXiv_2404_3117} has roughly twice the number density as the LRG sample used here. Given the number density trends in the BGS sample, we would likely find an even greater variance reduction for a higher-density LRG sample. Overall, it makes sense that our proposed method would lead to a larger variance reduction, especially on small scales, given the use of accurate gravity solvers and a more complex galaxy model. We leave detailed like-for-like comparisons of different control variate approaches to future work.

Finally, let us briefly comment on the qualitative trends shown in Fig.~\ref{fig:boost} to gain an understanding of how the results in this work would apply to other galaxy samples. The variance reduction depends on the cross-correlation between clustering measurements in the full-physics and gravity-only simulations. The clustering measures, in turn, are related to the number of galaxy pairs. Ultimately, the effectiveness of the CV method depends on how well the proxy model can predict variations of galaxy pairs. Our proxy model, a gravity-only simulation with a machine-learning model of the galaxy--halo connection, is able to predict variations in galaxy pairs because it has access to variations of the underlying cosmological volume by knowing what halos occupy it. However, the proxy model does not perfectly predict the number of galaxy pairs because it cannot perfectly predict how many galaxies each halo hosts, only an expectation value. Given these considerations, we expect the effectiveness of the CV method should improve with increasing galaxy bias, increasing number density, and increasing coefficient of determination, $R^2$, between predicted and expected galaxy occupation. Crucially, $R^2$ is much lower for ELGs, $R^2 \sim 0.2$, compared to other galaxy samples, $R^2 \sim 0.6 - 0.9$, explaining why it benefits the least from the CV method.

\subsection{Potential Applications}

As shown in Fig.~\ref{fig:boost}, the CV method demonstrated here can lead to significant increases in the predictive power for galaxy clustering, even on fairly non-linear scales. One prime application for this techniques would be the creation of mock galaxy clustering measurements from full-physics simulations. Full-physics simulations have the advantage of simulating all effects important for galaxy formation and galaxy and matter clustering such as the back-reaction of baryonic processes onto the matter distribution, also called baryonic feedback \citep{Leauthaud2017_MNRAS_467_3024, Lange2019_MNRAS_488_5771, Amodeo2021_PhRvD_103_3514, Amon2023_MNRAS_518_477}. By analyzing these mock measurements in the same way as the observational data, we can test whether inferences about cosmological parameters or the galaxy--halo connection, e.g. galaxy assembly bias, are robust \citep[see, e.g.,][]{Lange2023_MNRAS_520_5373, Beyond2ptCollaboration2024_arXiv_2405_2252}. By having more precise predictions, potential biases such as biases in cosmological parameters can be determined with higher precision. With the CV method, full-physics mock measurements can reach the galaxy density and effective volume of state-of-the-art observational data, ensuring the potential biases in the analysis method are not larger than statistical uncertainties.

Recently, the MilleniumTNG \citep{HernandezAguayo2023_MNRAS_524_2556} and FLAMINGO \citep{Schaye2023_MNRAS_526_4978} simulations have been introduced as the next-generation large-volume hydrodynamical simulations. Our proposed method is readily applicable to the FLAMINGO simulations. For example, the flagship {\sc L1\_m8} full-physics simulation ($V = 1^3 \, \mathrm{Gpc}^3$) has a gravity-only counterpart, {\sc L1\_m9\_DMO}, as well as a gravity-only simulation with the same resolution but much larger volume, {\sc L2p8\_m9\_DMO} ($V = 2.8^3 \, \mathrm{Gpc}^3$). Using the CV method, clustering predictions, especially for high-density samples, would rival the precision of predictions from the other full-physics flagship simulation, {\sc L2p8\_m9}, ($V = 2.8^3 \, \mathrm{Gpc}^3$) but with a factor of $10$ higher resolution. Contrary, the situation is slightly different for the MillenniumTNG simulation suite. MillenniumTNG has a gravity-only counterpart, {\sc MTNG740-DM}, of the flagship full-physics simulation, {\sc MTNG740}, ($V = 0.5^3 \, h^{-3} \, \mathrm{Gpc}^3$) as well as a large-volume gravity-only simulation, {\sc MTNG3000-DM-0.1$\nu$} ($V = 2.0^3 \, h^{-3} \, \mathrm{Gpc}^3$). However, the gravity-only simulations do not have the same underlying physics and resolution. Fortunately, a gravity-only simulation of the {\sc MTNG740} volume at the resolution of {\sc MTNG3000-DM-0.1$\nu$} only requires $2560^3$ matter and $640^3$ neutrino particles, i.e., a very modest computational effort.

\section{Conclusion}
\label{sec:conclusion}

State-of-the-art hydrodynamical simulations of galaxy formation simulate volumes significantly smaller than what can be observed with modern experiments such as DESI, limiting their usefulness in accurate galaxy clustering predictions. In this work, we developed a method based on the concept of control variates to make high-precision galaxy clustering predictions from existing full-physics simulations. We train a machine learning algorithm to learn the galaxy--halo connection and then combine this algorithm with a computationally cheaper gravity-only simulation. Through the use of control variates, we reduce the variance of the galaxy clustering predictions from the full-physics simulation without affecting their expectation value. Our main findings are highlighted below:

\begin{enumerate}
    \item We show that neural networks can effectively learn the complex interplay between halo properties and galaxy occupation, as demonstrated in Fig.~\ref{fig:r2}. We find that the halo maximum circular velocity, $V_\mathrm{max}$, is the most important secondary halo property besides halo mass to predict galaxy occupation.
    
    \item Across different galaxy samples, we find that the large-scale clustering amplitude is underpredicted by a mass-only model, signaling the presence of galaxy assembly bias in IllustrisTNG, in agreement with earlier works \citep{Hadzhiyska2020_MNRAS_493_5506, Hadzhiyska2021_MNRAS_508_698, Delgado2022_MNRAS_515_2733}. As shown in Fig.~\ref{fig:clustering}, we find that halo environment is the most important secondary halo property driving assembly bias.

    \item Fig.~\ref{fig:boost} displays the effective volume increases for different galaxy samples that emulate tracers in the DESI cosmology survey. The variance reduction increased as scale increased but is significant even on highly non-linear scales, $r \sim 1 - 10 \, \hmpc$. We calculate a volume increase of up to a factor of around $100$ for high-density BGS samples, a volume increase of roughly a factor of $10$ for LRGs across all clustering measures, and an increase of a factor of less than $10$ for almost all measures of ELGs.
\end{enumerate}

Our method is easily applicable to state-of-the-art simulations such as MilleniumTNG and FLAMINGO to produce highly-precise clustering predictions from high-resolution hydrodynamical simulations. In the near future, we plan to create such mock measurements to stress-test our ability to determine the properties of dark matter and dark energy from galaxy clustering in the strongly non-linear regime.

\section*{Acknowledgments}

We thank Boryana Hadzhiyska and Nickolas Kokron for insightful comments that
improved this paper. We thank the IllustrisTNG collaboration for publicly releasing the TNG300 simulations \citep{Pillepich2018_MNRAS_475_648, Springel2018_MNRAS_475_676, Nelson2018_MNRAS_475_624, Naiman2018_MNRAS_477_1206, Marinacci2018_MNRAS_480_5113, Nelson2019_ComAC_6_2}. This work made use of the following software packages: {\sc Astropy} \citep{AstropyCollaboration2013_AA_558_33}, {\sc halotools} \citep{Hearin2017_AJ_154_190}, {\sc matplotlib} \citep{Hunter2007_CSE_9_90}, {\sc NumPy} \citep{vanderWalt2011_CSE_13_22}, {\sc SciPy} \citep{Virtanen2020_NatMe_17_261}, {\sc scikit-learn} \citep{Pedregosa2011_JMLR_12_2825}, {\sc Corrfunc} \citep{Sinha2020_MNRAS_491_3022}, and {\sc Spyder}.

This research was supported in part through computational resources and services provided by Advanced Research Computing (ARC), a division of Information and Technology Services (ITS) at the University of Michigan, Ann Arbor. JUL acknowledges support from the Leinweber Center for Theoretical Physics, NASA grant under contract 19-ATP19-0058, and DOE under contract DE-FG02-95ER40899.

\bibliography{bibliography}

\end{document}